\documentclass[conference]{IEEEtran}
\IEEEoverridecommandlockouts
\usepackage{cite}
\usepackage{amsmath,amssymb,amsfonts}
\usepackage{algorithmic}
\usepackage{graphicx}
\usepackage{textcomp}
\usepackage{xcolor}
\def\BibTeX{{\rm B\kern-.05em{\sc i\kern-.025em b}\kern-.08em
    T\kern-.1667em\lower.7ex\hbox{E}\kern-.125emX}}

\usepackage{framed}
\usepackage{csquotes}

\usepackage{adjustbox}
\usepackage{array}
\usepackage{booktabs}
\usepackage{multirow}

\usepackage[bookmarks=false]{hyperref}

\newcolumntype{R}[2]{%
	>{\adjustbox{angle=#1,lap=\width-(#2)}\bgroup}%
	l%
	<{\egroup}%
}
\newcommand*\rot{\multicolumn{1}{R{90}{1em}}}

\usepackage{graphicx}
\usepackage{subcaption}

\begin{document}

\title{Asheetoxy: A Taxonomy for Classifying Negative Spreadsheet-related Phenomena\\
}

\author{\IEEEauthorblockN{Daniel Kulesz}
\IEEEauthorblockA{\textit{Institute of Software Technology} \\
\textit{University of Stuttgart}\\
Stuttgart, Germany \\
daniel.kulesz@iste.uni-stuttgart.de}
\and
\IEEEauthorblockN{Stefan Wagner}
\IEEEauthorblockA{\textit{Institute of Software Technology} \\
\textit{University of Stuttgart}\\
Stuttgart, Germany \\
stefan.wagner@iste.uni-stuttgart.de}
}

\maketitle

\begin{abstract}

Spreadsheets (sometimes also called Excel programs) are powerful tools which play a business-critical role in many organizations.
However, due to faulty spreadsheets many bad decisions have been taken in recent years.
Since then, a number of researchers have been studying spreadsheet errors.
However, one issue that hinders discussion among researchers and professionals is the lack of a commonly accepted taxonomy.

Albeit a number of taxonomies for spreadsheet errors have been proposed in previous work, a major issue is that they use the term error that itself is already ambiguous.
Furthermore, to apply most existing taxonomies, detailed knowledge about the underlying process and knowledge about the \enquote{brain state} of the acting spreadsheet users is required.
Due to these limitations, known error-like phenomena in freely available spreadsheet corpora cannot be classified with these taxonomies.

We propose Asheetoxy, a simple and phenomenon-oriented taxonomy that avoids the problematic term error altogether.
An initial study with 7 participants indicates that even non-spreadsheet researchers similarly classify real-world spreadsheet phenomena using Asheetoxy.

\end{abstract}


\section{Introduction}

For decades, spreadsheets have been business-critical tools used by companies, associations, schools and other organizations in all industrialized countries.
While spreadsheets offer many advantages, several cases show that the use of faulty spreadsheets can badly impact decisions and lead to loss of profit and reputation \cite{powell2009impact}.
Over the years, many researchers and professionals have been studying spreadsheet errors.
In this course, dozens of taxonomies for classifying spreadsheet errors have been developed and proposed.
However, there is still no commonly accepted taxonomy for spreadsheet errors, encumbering discussion among experts in the field.

\subsection{Problem Statement}

Researchers and professionals need a commonly acceptable taxonomy for classifying spreadsheet errors that is applicable to a wide number of contexts.

\subsection{Research Objective}

The primary goal of this study was to investigate the issues of existing spreadsheet error taxonomies and to enhance an existing taxonomy or develop a new one, providing a candidate capable of becoming a commonly accepted taxonomy.




\subsection{Context}

In previous work, we developed a framework for specifying, executing and analyzing tests in spreadsheets \cite{kulesz2014integrating}.
Later, we developed an approach named \enquote{Spreadsheet Guardian} \cite{kulesz2018spreadsheet} for protecting spreadsheet users from the introduction of faults in collaborative settings.
However, due to the lack of an applicable taxonomy, the discussion of related work was a challenge to us in these and other previous works.
We tried to overcome this by providing and using a taxonomy of our own.
However, we never felt truly comfortable with it.

In the course of writing a comprehensive work that includes a broad discussion of the state of the art in spreadsheet error research (not published yet), the first author realized that the previously used (unnamed) taxonomy required a major overhaul.
The result is the contribution at hand - a taxonomy named Asheetoxy that is a major revision of the previously unnamed taxonomy, accompanied by its first independent evaluation.

\section{Spreadsheet Error Taxonomies}

Before introducing our own taxonomy, we cover existing taxonomies for spreadsheet errors and criticism about them from previous work.
Following that, we argue why we think that previous attempts were unsuccessful and why we want to approach the problem differently with our taxonomy.

\subsection{Classification attempts}

\begin{figure}[htbp]
	\centering
	\includegraphics[width=0.49\textwidth]{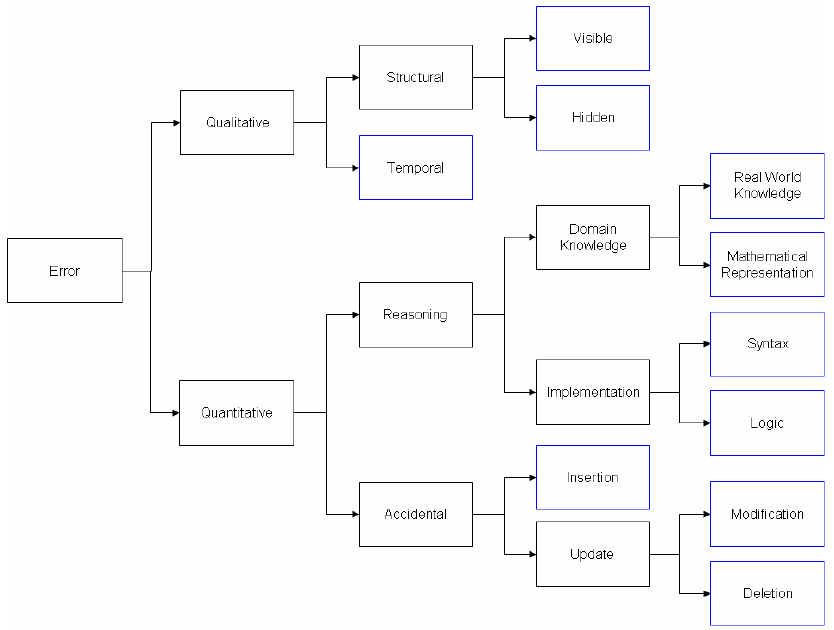}
	\caption{Revision by Purser and Chadwick of the Taxonomy by Rajalingham et al. (Figure taken from \cite{purser2006does})}
	\label{fig:taxo_purser_chadwick}
\end{figure}


Following early reports about spreadsheet errors \cite{brown1987experimental,cragg1993spreadsheet,ronen1989spreadsheet}, first attempts to develop a spreadsheet error taxonomy were undertaken \cite{galletta1993empirical,saariluoma1994transforming,panko1996spreadsheets,teo1997quantitative,rajalingham2000classification,rajalingham2005revised,howe2006factors}.
From these attempts, two taxonomies emerged that we would rate as most significant today: The taxonomy by Purser and Chadwick \cite{purser2006does} (a simplified optimization of the revised taxonomy by Rajalingham \cite{rajalingham2005revised}) and the revised taxonomy by Panko and Aurigemma \cite{panko2010revising} (a revision of the original taxonomy by Panko and Halveson \cite{panko1996spreadsheets}).

The taxonomy by Purser and Chadwick is shown in Figure \ref{fig:taxo_purser_chadwick}.
It is based on the idea originating from the first Taxonomy by Panko and Halverson \cite{panko1996spreadsheets} to distinguish between \enquote{qualitative} and \enquote{quantitative} errors.
According to the definition, quantitative errors are numeric errors that lead to wrong results.
On the other hand, qualitative errors are issues that \emph{only} lower overall quality of a spreadsheet and increase the risk of quantitative errors during future use.
One level below, the taxonomy further subdivides qualitative errors into structural errors (errors that have a permanent negative effect inside the spreadsheet) and temporal errors.
Regarding quantitative errors, the taxonomy distinguishes between accidental errors (the user had the right intention but did the wrong action) and reasoning errors (the user did not have a proper understanding and thus had the wrong intention and committed the wrong action).

\begin{figure}[hbp]
	\centering
	\includegraphics[width=0.49\textwidth]{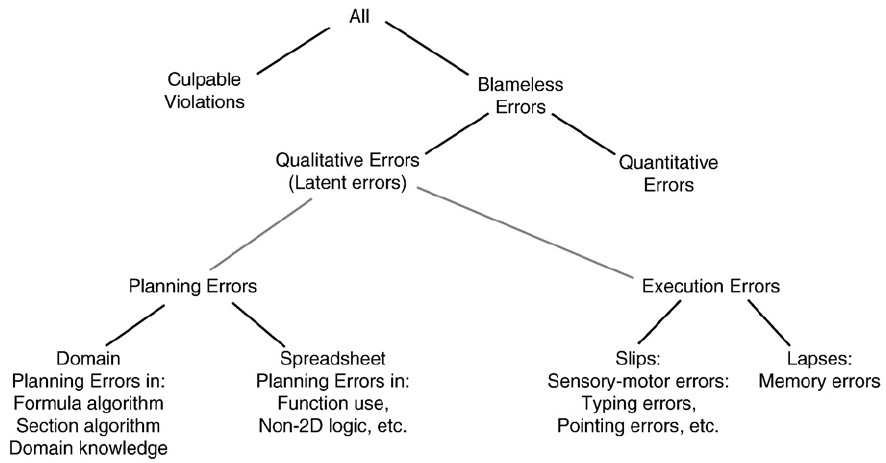}
	\caption{The taxonomy by Panko and Aurigemma \cite{panko2010revising}, a revision of the taxonomy by Panko and Halverson}
	\label{fig:taxo_panko_aurigemma}
\end{figure}

The taxonomy by Panko and Aurigemma (see Figure \ref{fig:taxo_panko_aurigemma}) distinguishes between culpable violations (errors done on purpose) and blameless errors (errors without fraudulent purpose) at the top level.
However, culpable violations (meanwhile several such cases are known, including manipulated transactions in the financial sector \cite{butler2009role} or counterfeit inspections of a nuclear plant \cite{thorne2012misuse}) are not distinguished any further.
Instead, the taxonomy focuses on errors where it can be assumed that the spreadsheet user committed a wrong action in good faith.
These blameless errors are then distinguished using the notion of qualitative and quantitative errors.
While quantitative errors are not further distinguished (the preceding taxonomy by Panko and Halverson \cite{panko1996spreadsheets} did that), the taxonomy subdivides qualitative errors into planning and execution errors.
This separation can be seen as an analogy with the concept of verification and validation \cite{wallace1989software} that is known from Software Engineering.
On the lower levels, the taxonomy further differentiates causes of the planning and execution errors.

\subsection{Criticism}

Powell et al. \cite{powell2008critical} state that there are several possibilities of distinguishing spreadsheet errors (i.e. based on cause, effect, form, stage or risk).
They argue that a commonly accepted taxonomy is hard to achieve as taxonomies are always developed for a particular purpose and context.
Apart from pointing out that there are no classifications that compare errors by development stage, Powell et al. criticize that existing taxonomies suffer from the following flaws:

\begin{itemize}
	\item They are defined imprecisely.
	\item They do not state the purpose and context for which they are intended explicitly.
	\item They do not include sufficient examples of errors that satisfy their categories.
	\item It was not demonstrated that different users choose consistent categories for classifying errors using the taxonomies.
	\item The categories they use are too vague (pointing directly to the distinction between qualitative and quantitative errors).
\end{itemize}

The taxonomy discussed by Panko and Aurigemma \cite{panko2010revising} can be seen as a direct reaction to this critism.
Panko and Aurigemma stated that they agree that the distinction between qualitative and quantitative errors is not always clear.
This can be illustrated using the corner-case of a formula that is overwritten with a constant value.
One could argue that the result of the spreadsheet would be correct as long as the input values remain untouched.
Therefore, the overwritten formula could be seen as a qualitative error because the bottom-line value of the spreadsheet is correct.
On the other hand, one could argue that there is a quantitative error because once the input values are changed, the bottom-line value can get wrong.
Furthermore, one could argue that it is a qualitative error because the overwritten formula has become an input cell, but the issue is that this fact is not immediately visible.

A different concern raised by Panko and Aurigemma is the issue that most taxonomies are theory-informed and not phenomena-oriented.
The issue with theory-informed taxonomies is that errors cannot be categorized using them if their creation process cannot be analyzed in-depth.

We agree with Powell et al.'s criticism as well as with the concerns raised by Panko and Aurigemma in all essential points.
From the classification discussion, we further derive that it is crucial for a taxonomy to define whether or not it distinguishes between the actual and possible future states of a spreadsheet.
Overall, we could not agree more that the absence of phenomena-oriented taxonomies is a major issue: especially, since over the last few years a number of valuable corpora with known faulty spreadsheets have become available (such as  \cite{cunha2012towards},\cite{Hofer2013},\cite{Ausserlechner2013},\cite{getzner2015improvements},\cite{cheung2016custodes},\cite{schmitz2016finding} or \cite{GetznerHW17}).
However, for most of the spreadsheets in these corpora, it may be impossible to precisely reconstruct and understand \emph{why} an error was introduced, making the proper classification of the error using a theory-informed taxonomy virtually impossible.
Therefore, it is not a big surprise that the authors of the accompanying publications did not use a common taxonomy when describing their observations or concluding their findings.

\subsection{Ambiguity of the term error}

Apart from the concerns raised both by Powell et al. and Panko and Aurigemma, we think that there is even a deeper problem with the classification of spreadsheet errors: the term error itself is simply too ambiguous.
The huge number of possibilities for distinguishing between types of spreadsheet errors at least seems like an indicator for this ambiguity.
Exactly the same issue is well known in the field of software engineering, where the IEEE Standard Glossary of Software Engineering Methodology (IEEE Std. 610-12-1990) gives four substantially different meanings for the term error:

\begin{itemize}
	\item \enquote{The difference between a computed, observed, or measured value or condition and the true, specified, or theoretically correct value or condition.
		For example, a difference of 30 meters between a computed result and the correct result.}
	\item \enquote{An incorrect step, process or data definition. For example, an incorrect instruction in a computer program.}
	\item \enquote{An incorrect result. For example, a computed result of 12 when the correct result is 10.}
	\item \enquote{A human action that produces an incorrect result. For example, an action on the part of a programmer or operator.}
\end{itemize}

\begin{figure}[htbp]
	\centering
	\includegraphics[width=0.49\textwidth]{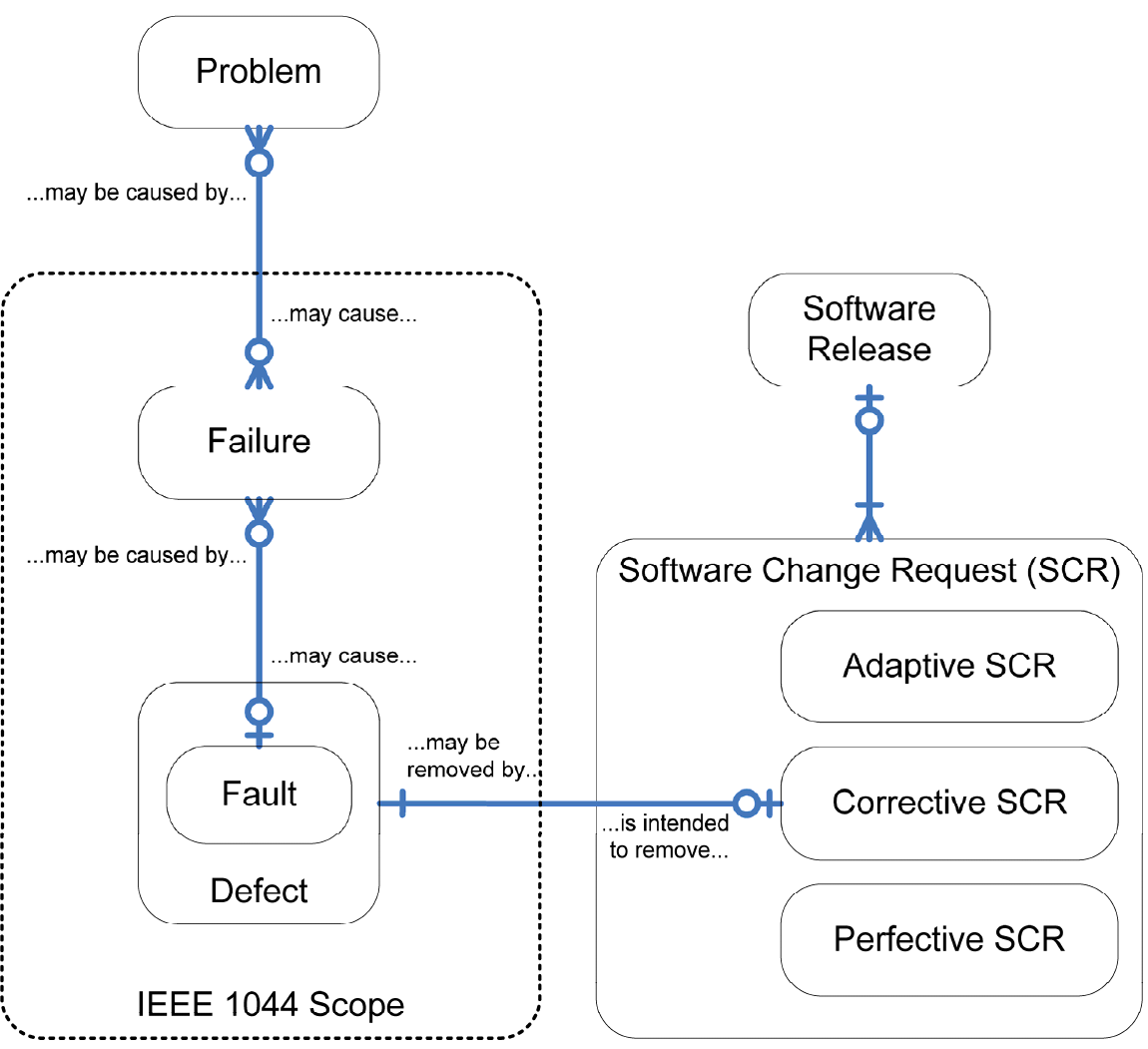}
	\caption{Relation between software anomalies and maintenance (Figure extracted from IEEE Std. 1044-2009)}
	\label{fig:grafik1044}
\end{figure}

In the software engineering field, there are two influential taxonomies that deal with this issue: the IEEE standard that models the relation between software anomalies and maintenance activities (IEEE-Std. 1044:2009) as illustrated in Figure \ref{fig:grafik1044} and the set-theory based model proposed by Liggesmeyer \cite{liggesmeyer2009software}.

In the end, we raised the following question: As there are viable alternatives for the term error in the software engineering world, can the issue be solved for spreadsheets likewise?
This led us to the conclusion that, due to its striking ambiguity, the use of the term error should be completely avoided in the context of spreadsheets.
While one could certainly define one of the meanings defined in IEEE Std. 610-12-1990 as the \emph{correct} one (e.g. only unintentionally wrong human actions could be named errors), we think that this could be counter-intuitive for anyone already used to a different understanding of this problematic term.

\section{Asheetoxy}

To address the issues discussed in the previous sections, we developed the phenomenon-oriented taxonomy named Asheetoxy.
Consequently, Asheetoxy does not use the term error at all.
Asheetoxy was influenced by the IEEE Std. 1044-2009 and Liggesmeyer's terms but it also borrows ideas from the original Taxonomy by Panko and Halverson.
Yet, Asheetoxy is not a cherry-pick-remix of existing taxonomies but rather a recomposition with added elements and distinctions that are directly targeted at the particularities of spreadsheets.

\begin{figure}[htpb]
	\centering
	\includegraphics[width=0.5\textwidth]{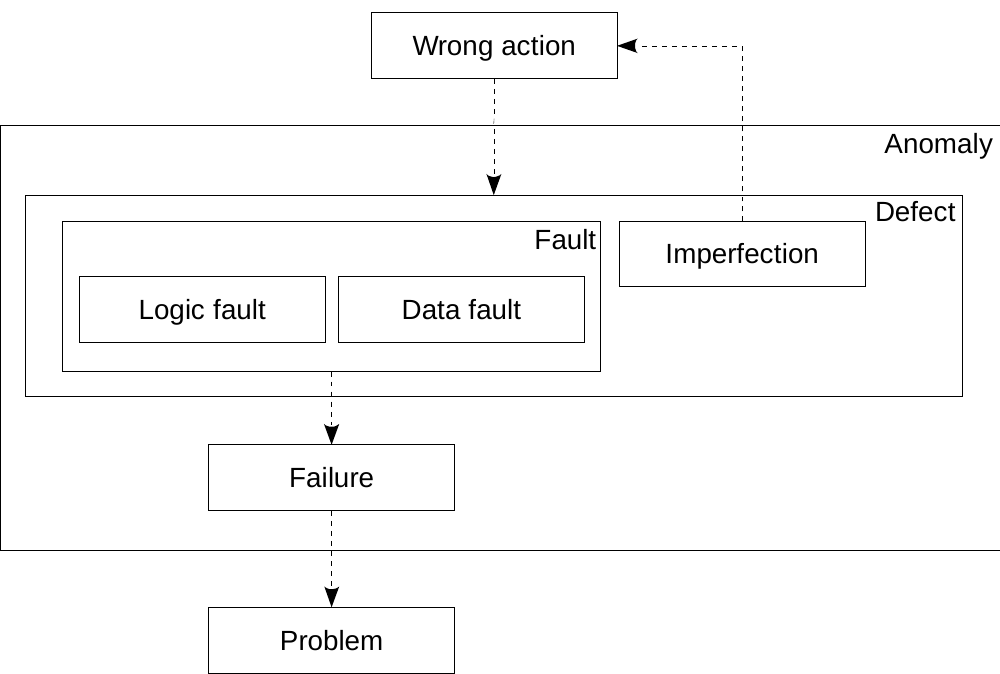}
	\caption{Terms in Asheetoxy}
	\label{fig:taxo3redraw}
\end{figure}

An overview of Asheetoxy with all of its terms and relations is provided in Figure \ref{fig:taxo3redraw}. 
The arrows between the terms state a \emph{can cause} relationship (e.g. a wrong action can cause a defect, a defect can cause a failure, etc.).

\begin{figure*}[hbpt]
	\centering
	\includegraphics[width=0.4\textwidth]{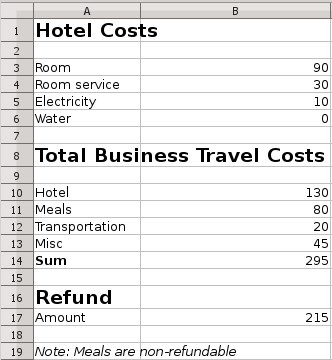}
	\hspace{1.5cm}
	\includegraphics[width=0.4\textwidth]{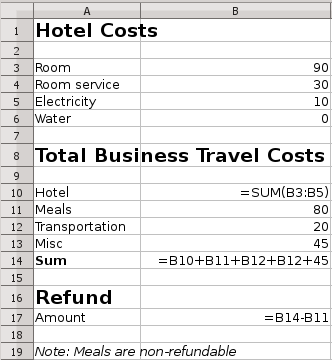}
	\caption{Spreadsheet with anomalies in normal view (left) and in formula view (right)}
	\label{fig:anomalyexample}
\end{figure*}

In the following, we provide definitions for the terms used.
To ease understanding, we use the example spreadsheet illustrated in Figure \ref{fig:anomalyexample} (on the next page).

\subsection{Anomaly}

\leftbar
The term \enquote{anomaly} refers to anything that is potentially negative and that can be found in a given spreadsheet.
\endleftbar

Effects that occur in the real world due to a \emph{bad} spreadsheet (e.g. someone is refused a loan because the spreadsheet showed wrong numbers) but outside of the spreadsheet itself are not regarded as spreadsheet anomalies by Asheetoxy.
This is because such effects can occur even if no spreadsheets were involved in the process.
Often, it is not transparent to outsiders whether or not spreadsheets were used anyway.

\subsection{Wrong action}

\leftbar
The term \enquote{wrong action} describes human actions during the use of spreadsheets that have negative effects.
\endleftbar

In our example (see Figure \ref{fig:anomalyexample}), when constructing the sum formula in cell B10, the spreadsheet user might have selected an incomplete cell area (B3:B5 instead of B3:B6).
In the given example, there may be at least three possible explanations for the wrong action taken by the user:

\begin{itemize}
	\item The user selected the wrong area by mistake. Although he knew that the costs of water should be included in the hotel costs, he simply clicked in the wrong cell.
	\item The user selected the wrong area because he thought that the cost of water \emph{does not} belong to the hotel costs.
	\item The user selected the wrong area intentionally. He knew very well that the cost of water does not belong to the hotel costs but wanted the spreadsheet to calculate a lower refund on purpose. To hide this fraudulent intent he wanted to make it look like an unintentional mistake.
\end{itemize}

However, an outside observer can only notice the fact \emph{that} the user took a wrong action but not \emph{why}.
As Asheetoxy is purely phenomenon-oriented, it regards \emph{all} of these cases as wrong actions and does not distinguish between them any further.

\subsection{Defect}

\leftbar
A spreadsheet contains a \enquote{defect} if it has undesired data, formulas or formats in its current state.
\endleftbar

When reading, modifying or inspecting a spreadsheet, a spreadsheet user might notice cells that he suspects of containing a defect.
This might be the case with the previously discussed formula in cell B10.
But also the sum formula in cell B14 could be classified as having a defect because it refers to cell B12 twice.
Another good candidate for a defect in this formula is the fact that, instead of referring to cell B13, the formula only sums up B13's value (45) as a hardcoded constant.

Depending on the potential effect of a defect, it can be further classified as a fault or an imperfection.

\subsection{Imperfection}

\leftbar
An \enquote{imperfection} is a defect with a potentially \emph{qualitative} impact - this means that it cannot have direct negative effects on the correctness of a spreadsheet.
\endleftbar

Typically, imperfections influence non-functional aspects such as usability or maintainability.
Therefore, the presence of imperfections increases the likelihood of wrong actions in the future.

In the example above, one could argue that it is not immediately clear whether or not the formula in cell B14 intentionally summed up the value of B12 twice.
It could be that it is perfectly correct to sum up B12 twice because it is supposed that transportation costs are expected to be provided for one way only (but are refunded for both ways).
However, in this case the label in cell A12 would be an imperfection as it is not clear.

One could also argue that for future maintenance, the intention of summing up B12 twice would be more clear if the formula in cell B14 made this explicit by using a multiplication operator (=B10 + B11 + 2 * B12 + B13).
Also, the use of the plus operators instead of a sum function could be rated as an imperfection.
However, in this example it would be questionable whether the resulting formula (=SUM(B10:B11)+2*B12+B13) would be easier to understand.

\subsection{Fault}

\leftbar
A \enquote{fault} is a defect with a potentially \emph{quantitative} impact - this means that it can have a direct negative effect on the correctness of a spreadsheet.
\endleftbar

Asheetoxy distinguishes between two types of faults: logic faults and data faults.

\subsubsection{Logic Fault}

\leftbar
A \enquote{logic fault} is present in a spreadsheet if the spreadsheet contains a formula that might compute wrong results for correct input values.
\endleftbar

If we assume that in our example the costs of water should be included in the total hotel costs, there would be a logic fault in cell B10.
Even if cell B6 has the value 0, the logic fault is still there because if a spreadsheet user entered a non-zero value in cell B6, the result in cell B10 (and the total result in cell B17) would be wrong.

\subsubsection{Data Fault}

\leftbar
A \enquote{data fault} is present in a spreadsheet if the spreadsheet contains data or formats due to which it might compute wrong results.
\endleftbar

Even if a spreadsheet is free from logic faults, it might still compute wrong results due to data faults.
Given the assumption that the actual costs of room service are not 30 but 50, there would be a data fault in cell B4.
However, this effect generally known as \enquote{Garbage In, Garbage out} may not necessarily occur:
Let's assume that the costs of misc would be 45, but due to a wrong action a spreadsheet user entered the value 450 in cell B13.
Despite the data fault in cell B13, the total result in cell B17 would still be correct.
Therefore, data faults \emph{may} lead to wrong results but they do not necessarily have to.

\subsection{Failure}

\leftbar
A \enquote{failure} is present in a spreadsheet if the spreadsheet contains a fault that has an actual (and not just a potential) effect.
\endleftbar

In the given example, there would be a failure in cell B10 if the costs of water should be included in the total hotel costs and cell B6 contained a non-zero value (e.g. 5).
It would be caused by the logic fault in cell B10 discussed previously.
Due to the failure in cell B10, there would also be a failure in cells B14 and B17 as these depend on B10.

Even if a spreadsheet contains multiple faults, it is not guaranteed that failures will be visible.
In some cases, faults can even neutralize each other and hide failures.
Let's assume the following scenario:
A spreadsheet user correctly inserted the value 40 in cell B6.
Due to the logic fault in cell B10's formula, the computed value in B10 is wrong (it is too low by 40).
Thus, there is a failure in cell B10.
But due to the imperfection in cell A12, the spreadsheet user thought that the value in cell B12 is for a two-way travel.
Therefore, he committed a wrong action and inserted the value 40 in cell B12 (while 20 would be the correct way for a one-way travel).
Thus, there is a data fault in cell B12.
Overall, the spreadsheet now contains two faults (a logic fault in cell B10 and a data fault in cell B12) and one failure (also in cell B10).
Yet, the sums in cell B14 and B17 are still correct (no observable failures here!) because, in this case, due to the particular values the two faults neutralize each other.

\subsection{Problem}

\leftbar
We use the term \enquote{problem} if a failure in a spreadsheet leads to direct negative consequences in the real world.
\endleftbar

In the given example, there would be a problem if the spreadsheet had been used for years to issue refunds, resulting in incorrect refunds for some travellers.
In this case, wrong amounts of money would have been transferred to the travellers.
Apart from the financial damage, the organization using the spreadsheet could also suffer reputational damage.
The damage probably would be even greater if it turned out that the wrong actions taken in the first place were not accidents but originated from fraud.

\section{Evaluation Planning}

As proposed by Powell et al., we designed a study to find out whether different users consistently classify spreadsheet-related phenomena when applying Asheetoxy.
However, we regarded Asheetoxy in its current state as a late draft, not as a finished taxonomy.
Therefore, the primary focus of our evaluation was to discover open issues in order to tackle them in the next revision.
Thus, we covered questions such as (1) whether or not Asheetoxy is precisely defined or (2) whether or not Asheetoxy provides sufficient examples only briefly.
For the same reason, we did not evaluate whether or not classifications issued by other users differ from those we would have chosen.

\subsection{Goal}
\label{sec:goal}

We defined the following experimental goal (style as proposed by Jedlitschka and others \cite{jedlitschka2008reporting}):

\begin{itemize}
	\item Goal: Analyze researchers classifying spreadsheet-related phenomena using Asheetoxy
	
	For the purpose of evaluating the consistency of their ratings
	
	With respect to the terms chosen for each phenomenon
	
\end{itemize}

\subsection{Participants}

The 7 (male) participants in our study were researchers at our computer science faculty and friends or co-workers of the first author.
All of them had a software engineering background.
They were roughly between 25 and 40 years old (we did not gather precise data for privacy reasons).
The participants were acquired by issuing a call for participation in an electronic chat room and \enquote{direct marketing} in the university building.

\subsection{Procedure}

The experiment was designed to be taken both off-site and without any help by the experimenter.
The participants received the experimental material as a PDF file and were asked to print it out before using it.
They were also asked to provide their results by filling out an online questionnaire (self-hosted using the free \enquote{Limesurvey} software package).

\subsection{Experimental Material}

The experimental material\footnote{available to researchers from https://doi.org/10.5281/zenodo.1406000} included three parts: (1) A page describing the background and the actual tasks, (2) the description of Asheetoxy as presented in section III, and, (3) three pages on which 9 spreadsheet-related phenomena were described.
All descriptions of spreadsheet-related phenomena were taken from the \enquote{horror stories} section of the EuSpRIG website \cite{horror}.
They were excerpts of original press articles that reported on spectacular problems resulting from spreadsheet failures.
However, most of the articles focused on the problem itself and did not go into detail about the originating wrong action or the actual anomaly.
In each of the excerpts, we drew one or two boxes (12 in total) around a few words that we found descriptive of a particular phenomenon.

\subsection{Tasks}

The experiment was designed for a total duration of 35 minutes (this expected duration was stated on the paper but not controlled).
The participants were asked to complete three tasks:

\begin{enumerate}
	\item read the description of Asheetoxy
	\item read the excerpts about the spreadsheet-related phenomena and, for each box in each excerpt, choose the term from Asheetoxy that they find most appropriate
	\item fill out a short questionnaire
\end{enumerate}

\subsection{Hypotheses}

To target our goal (see section \ref{sec:goal}), we formulated hypothesis $H_1$:

\begin{itemize}
	
	\item[$H_1$] For each phenomenon, the majority of the participants classify it consistently.
	\item[$H_{1_0}$] For each phenomenon, the majority of the participants do not classify it consistently.
	
\end{itemize}

Apart from looking for exact matches, we also wanted to check the compatibility of classifications.

\subsection{Analysis Procedure}

The analysis covered the participants' classifications and the feedback they provided in the questionnaire.

\section{Execution and Analysis}

The experiment was executed as planned with the small deviation that we also requested personal feedback from the participants in a quick interview after they finished the experiment.

In the following, we present the gathered data.

\subsection{Descriptive Statistics}

The distribution of classifications for the particular phenomena is provided in Table \ref{tab:results}.
	
	\begin{table}[h] \centering
		\begin{tabular}{clccccccccc|ccc}
			& & \multicolumn{12}{c}{Asheetoxy terms} \\
			& & \rot{Wrong action} & \rot{Anomaly} & \rot{Defect} & \rot{Fault} 
			& \rot{Data fault} & \rot{Logic fault} & \rot{Imperfection} 
			& \rot{Failure} & \rot{Problem} & \rot{more than one} &  \rot{No matching term} & \rot{Not sure at all}\\
			\midrule
			& Ph. 1             & - & -  & -  &  - &  - & 1 & 1 & 3 & 1 & 1 & - & - \\
			& Ph. 2             & 2 & -  & -  & -  & -  & - & - & - & 5 & - & - & - \\
			& Ph. 3             & 5 & -  & -  & -  & 2  & - & - & - & - & - & - & - \\
			& Ph. 4             & 4 & -  & -  & -  & -  & 2 & 1 & - & - & - & - & - \\
			& Ph. 5             & - & -  & -  & -  & 1  & - & - & 1 & 1 & 1 & 3 & - \\
			& Ph. 6             & - & 3  & -  & -  & -  & - & - & 2 & 1 & 1 & - & - \\
			& Ph. 7             & - & -  & -  & -  & -  & 1 & - & - & 5 & - & 1 & - \\
			& Ph. 8             & - & -  & -  & -  & -  & - & - & - & 5 & - & 2 & - \\
			& Ph. 9             & - & 1  & -  & -  & -  & 1 & - & 2 & 1 & 2 & - & - \\
			& Ph. 10            & - & -  & -  & -  & 4  & 2 & - & - & - & 1 & - & - \\			
			& Ph. 11            & - & -  & -  & -  & -  & 4 & 1 & 1 & - & 1 & - & - \\			
			& Ph. 12            & - & 2  & -  & -  & -  & - & 1 & 1 & - & 2 & 1 & - \\
			
			\bottomrule
		\end{tabular}
	
		\caption{Classification results}
		\label{tab:results}
	\end{table}

\begin{table}[h] \centering
	\begin{tabular}{llcc}
		& Largest &  size & \# of other distinct\\
		& compatible set & & (incompatible) sets\\
		\midrule
		Ph. 1 & failure & 4 & 3              \\
		Ph. 2 & problem  & 5 & 1              \\
		Ph. 3 & wrong action  & 5 & 1              \\
		Ph. 4 & wrong action  & 4 & 2              \\
		Ph. 5 & no match  & 3 & 3              \\
		Ph. 6 & anomaly  & 6 & 1              \\		
		Ph. 7 & problem  & 5 & 2              \\		
		Ph. 8 & problem  & 5 & 1              \\
		Ph. 9 & anomaly  & 4 & 3              \\		
		Ph. 10 & data fault  & 4 & 2              \\		
		Ph. 11 & logic fault  & 4 & 2              \\						
		Ph. 12 & anomaly  & 4 & 3              \\				
		
		\bottomrule
	\end{tabular}
	
	\caption{Compatibility of classifications}
	\label{tab:results2}
\end{table}

To check the compatibility of classifications, we analyzed whether or not they are contained in a compatible set (e.g. a logic fault and a defect are in a compatible set because a logic fault is just a special type of a defect; on the other hand, e.g., a wrong action and an imperfection are in distinct sets).
To get a comprehensive picture, we checked how many classifications were contained in the largest compatible set and how many incompatible sets existed.
If respondents answered that more than one term was appropriate, we counted the answer as compatible only if one of the terms matched the largest compatible set.
Table \ref{tab:results2} provides the results that were obtained by detailed analysis of the responses for each phenomenon.

\subsection{Hypothesis examination}

The hypotheses were examined based on the consistency metric of compatible sets (Table \ref{tab:results2}).
Depending on the phenomenon, the size of the largest compatible set (i.e. the number of all compatible terms) differed between 3 and 6 with a median of 4 and an average of 4.42.
Overall, these numbers support $H_1$ for all phenomena except phenomenon 5.

None of the phenomena (including phenomenon 5) supports $H_{1_0}$.
Hence, we reject $H_{1_0}$.

\subsection{Subjective impressions}

\begin{figure}[hbp]
	
	\centering
	
	\begin{subfigure}[b]{0.49\textwidth}
		\includegraphics[width=1\textwidth]{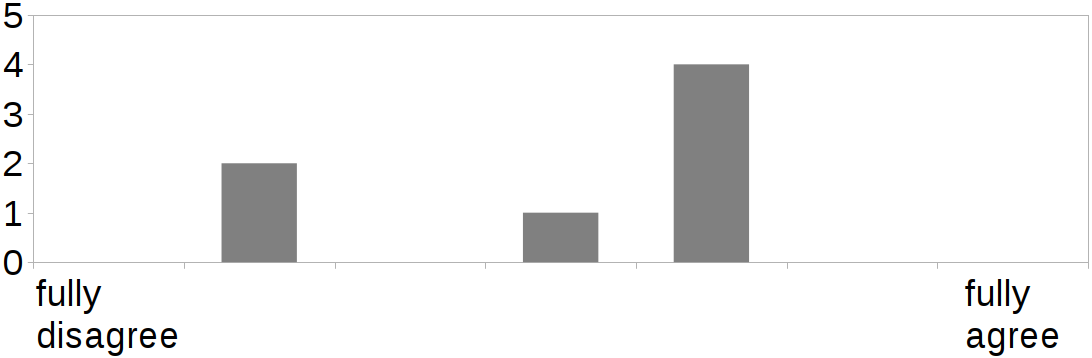}
		\caption{Q: It was easy to find matching terms for the examples in the news articles.}
	\end{subfigure}
	
	\vspace{0.5cm}
	
	\begin{subfigure}[b]{0.49\textwidth}
		\includegraphics[width=1\textwidth]{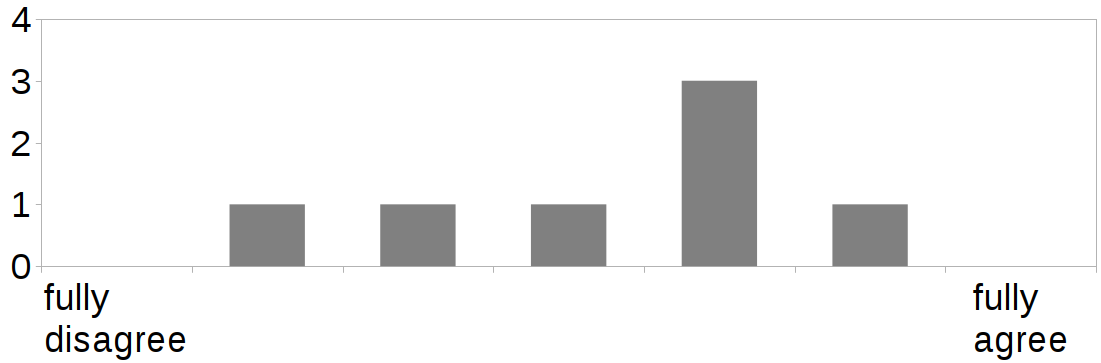}
		\caption{Q: Most of the news articles provided enough detail to find the proper term for the particular phenomenon.}
	\end{subfigure}
	
	\caption{Participants' answers to task-related questions}
	\label{fig:application-experience}
\end{figure}

\begin{figure}[tpbh]
	\centering
	
	\begin{subfigure}[b]{0.49\textwidth}
		\includegraphics[width=1\textwidth]{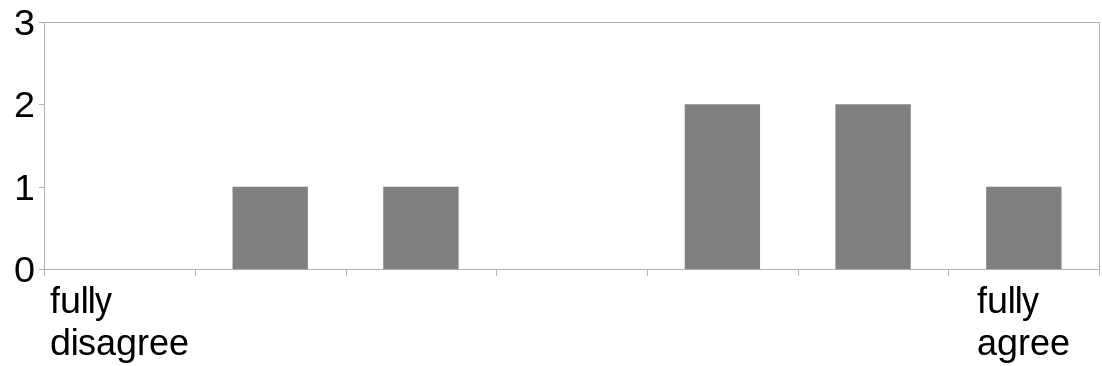}
		\caption{Q: The terms used in Asheetoxy are intuitive.}
	\end{subfigure}
	
	\vspace{0.5cm}

	\begin{subfigure}[b]{0.49\textwidth}
		\includegraphics[width=1\textwidth]{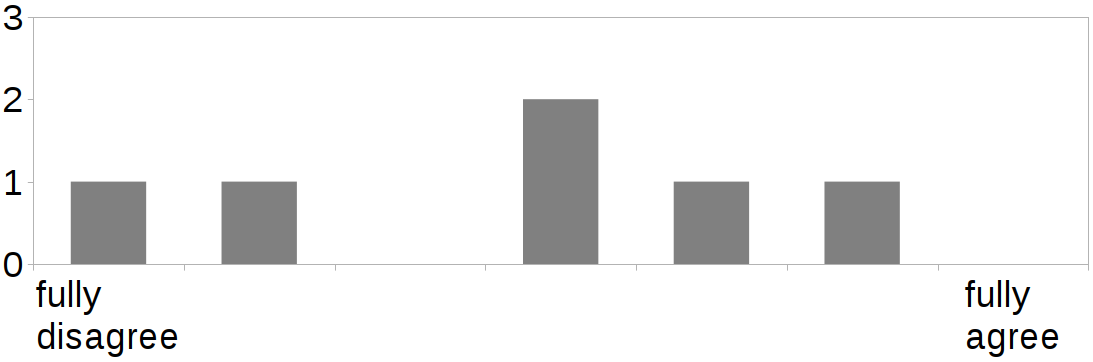}
		\caption{Q: I would replace some terms in Asheetoxy by other terms.}
	\end{subfigure}

	\vspace{0.5cm}
	
	\begin{subfigure}[b]{0.49\textwidth}
		\includegraphics[width=1\textwidth]{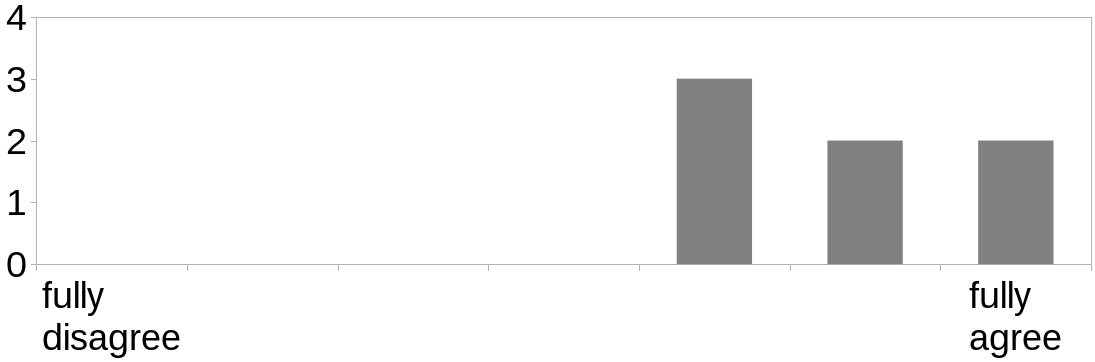}
		\caption{Q: Overall, the paper describing Ahseetoxy was easy to understand.}
	\end{subfigure}
	
	\vspace{0.5cm}
	
	\begin{subfigure}[b]{0.49\textwidth}
		\includegraphics[width=1\textwidth]{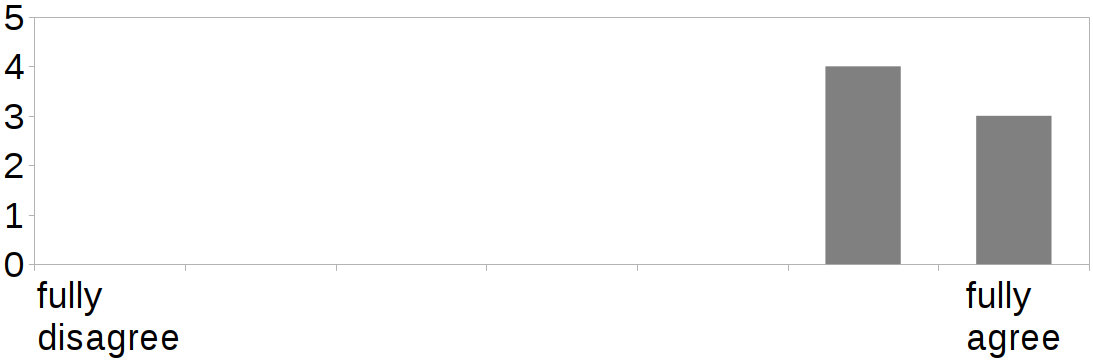}
		\caption{Q: The example used in the Asheetoxy paper supported me in understanding it.}
	\end{subfigure}

	\caption{Participants' answers to Asheetoxy-related questions}
	\label{fig:asheetoxy-impressions}
\end{figure}

In the questionnaire, the participants provided answers regarding their subjective impression about understanding Asheetoxy (Figure \ref{fig:asheetoxy-impressions}) as well as applying it to the given news article excerpts (Figure \ref{fig:application-experience}).

The data indicate that most participants found the description of Asheetoxy easy to understand and its example helpful.
However, several participants had reservations regarding the particular terms used in Asheetoxy.
Furthermore, the application of Asheetoxy's terms to the particular phenomena was challenging for some participants and most participants were not satisfied with the phenomenon-classification-relevant level of detail provided in the news article excerpts.

\section{Discussion}

\subsection{Evaluation of Results and Implications}

The results suggest that classifications made by different researchers using Asheetoxy are mostly consistent.
Surprisingly, no participant used the classes \enquote{defect} and \enquote{fault}.
The results also indicate that the current description of Asheetoxy (and its example) is adequate.
Also, at least some technical detail about a phenomenon seems to be beneficial in order to classify it.
This seems not to be the case with the majority of the news article excerpts used in our study.

\subsection{Threats to Validity}

Due to its pilot character, the study is subject to a number of threats to validity, including the following:

\begin{itemize}
	\item The sample size was tiny.
	\item The selected subjects may not be representative of real spreadsheet researchers or professionals engaged in negative spreadsheet phenomena.
	\item Due to the arbitrary selection we cannot assure that the level of detail about spreadsheet phenomena described in the selected article excerpts are representative of similar news articles.
	\item Although the classifications chosen by the participants were mostly consistent with each other, the evaluation did not shed any light on the question whether they were also consistent with the intent of the taxonomy's authors.
\end{itemize}

\subsection{Lessons Learned}

In the course of detaching Asheetoxy as an independent work and evaluating it with other researchers, we gained several invaluable insights.
It became apparent that there are effects outside of spreadsheet artifacts (namely wrong actions and problems). Thus, the term spreadsheet anomaly used by us before was too narrow (we find the broader notion of a \enquote{spreadsheet-related negative phenomenon} a better fit).
Also, our previous unnamed taxonomy missed this important separation between events that occur \emph{inside} a spreadsheet and events that occur \emph{outside} a spreadsheet (i.e. the spreadsheet's environment).

By actually carrying out the experiment on the news article excerpts and receiving feedback from the participants we also learned that it might also be a good idea for a taxonomy to explicitly state which information about the particular phenomenon is actually required for applying it.

\section{Conclusions and Future Work}

In this work, we proposed and successfully piloted a taxonomy that allows spreadsheet-related negative phenomena to be classified without using the problematic notion of a spreadsheet error.
It would be great if Asheetoxy aided others in discussing spreadsheet issues more precisely.

For future work, we think that it would make sense to first enhance some details in the description of the taxonomy and its terms that became apparent in the evaluation before running an evaluation on a broader scope.
However, we think that it could be interesting to classify more \enquote{tangible} anomalies in spreadsheet corpora rather than phenomena described in news articles.

\bibliography{bibliography}
\bibliographystyle{IEEEtran}

\end{document}